\documentclass{article}
\usepackage{epsf}
\usepackage{emulateapj}
\begin{document}
\lefthead{Robertson \& Leiter}
\righthead{How Black are BHC?}

\title{How Black are Black Hole Candidates?}

\author{Stanley L. Robertson\altaffilmark{1} and Darryl J. Leiter\altaffilmark{2}}
\altaffiltext{1}{Dept. of Physics, Southwestern Oklahoma State University,
Weatherford, OK 73096}
\altaffiltext{2}{FSTC, Charlottesville, VA 22901}

\begin{abstract}
In previous work we found that many of the spectral properties of
x-ray binaries, including both galactic black hole candiates (GBHC) and
neutron stars, were consistent with the existence of intrinsically
magnetized central objects. Here we review and extend the
observational evidence for the existence of intrinsically
magnetized GBHC and show that their existence is consistent with
a new class of solutions of the Einstein field equations of General Relativity.
These solutions are based on a strict adherence to the Principle of
Equivalence, which prevents the time-like geodesics of physical matter
from becoming null on trapped surfaces of infinite red shift.
The new solutions emerge from the fact that the structure and radiation
transfer properties of the energy-momentum tensor on the right hand side of
the Einstein field equations must have a form that is consistent with
this Principle of Equivalence requirement. In this context, we
show that the Einstein field equations allow the existence of highly
red shifted, magnetospheric, eternally collapsing objects (MECO) which
do not have trapped surfaces which lead to event horizons. Since MECO
lifetimes are many orders of magnitude greater than a Hubble time,
they provide an elegant and unified framework for understanding the
broad range of observations associated with
GBHC and active galactic nuclei.
\end{abstract}


\section{Introduction}
In previous work (Robertson \& Leiter 2002) we presented evidence
for the existence of intrinsic magnetic moments in galactic
black hole candidates (GBHC).
We proposed that this observational result was consistent
with the idea that the fundamental structure of General Relativity
allows the existence of eternally collapsing objects (ECO) without
event horizons (Mitra, 2000, 2002). Since the energy-momentum tensor
of the right hand side of the Einstein field equations serves as both
a source of curvature and a generator of equations of motion of matter,
any constraints on the latter will impact the former. In this regard,
the Principle of Equivalence (POE) requirement that the time-like
geodesics of matter cannot become null is a constraint that must be
encompassed by elements included in the energy-momentum tensor.
In Section 2 we show that strict adherence to the POE
implies that the Einstein field equations possess physical solutions
consistent with the existence of highly red shifted, magnetospheric eternally
collapsing objects (MECO) in which trapped surfaces of infinite redshift
leading to event horizons cannot form. Unlike the cold, catalyzed matter
of neutron stars, with only limited masses that can be supported by degeneracy
pressure, MECO are hot and radiatively supported while slowly collapsing. Since
the MECO solutions have observable lifetimes which
are many orders of magnitude greater than a Hubble time, their existence
as the central compact component of active galactic nuclei
(AGN) is compatible with the broad range of observations associated with
GBHC and AGN.

This work is intended as a first introduction to MECO phenomena. We hope to
make the existence of MECO plausible and show that they fit very comfortably
within the complex phenomenology of compact astrophysical objects, while
necessarily leaving many important and difficult
issues for later work. In Sections 3 and 4, we
first consider some general properties of quiescent and active MECO.
In Section 5 we examine the compatibility of MECO
models of GBHC and AGN with a variety of astrophysical
observations. We find that compatibility with observations requires that the
MECO rates of rotation be relatively slow.
We find that generally diamagnetic accreting plasma interacting
with the magnetic field of a MECO via an
accretion disk provides an elegant and unified framework for
understanding compact x-ray sources. In particular, the MECO model
accounts for the high state `ultrasoft' radiation (White \& Marshall 1984),
the high state power law emissions, the spectral state switch, including the
radio-loud and radio quiet states, low state jets and equatorial outflows,
and the quiescent luminosities of GBHC.

There is a plethora of piece meal black hole models of these
various phenomena. For example, comptonizing coronae near event horizons,
bulk flow comptonization, accretion disk coronae, and magnetic flares on
accretion disks have all been invoked to explain the hard spectral tail
of black hole GBHC. Radiatively inefficient advective flows at high
accretion rates have been proposed to explain their quiescent power-law
emissions while overlooking the obvious fact that such flows
do not occur for neutron star binaries. Since there is no unified black
hole model, while the MECO model provides a comprehensive and unified
approach, the replacement of black holes by MECO represents a paradigm
shift in the astrophysics of compact objects. To believe in
the existence of trapped surfaces and event
horizons is to believe that nature has provided
a way to accelerate particles with non-zero rest mass to exactly the
speed of light. The POE implies that this is impossible and the
issue can be settled by observations of magnetospheric phenomena. MECO
have magnetic moments, black holes do not.

\section{Principle of Equivalence and MECO}
In General Relativity the Principle of Equivalence (POE) requires
that Special Relativity must hold locally for all freely
falling time-like observers. This `medium-strong' form (Wheeler \& Ciufolini 1995)
of the principle can be expressed as a tensor relationship, which
means that in a general curved spacetime, physical matter
must follow time-like world lines such that the associated
invariant line interval obeys
\begin{equation}
ds^2 = g_{ij}dx^i dx^j = c^2d\tau^2 = \frac{c^2 dt^2} {(1 + z)^2} > 0
\end{equation}
where (1 + z) is the red shift given by
\begin{equation}
(1 + z) = \frac{dt}{d\tau}
\end{equation}
The POE requires that the proper time $d\tau$
along time-like world lines of physical matter
must exist and world lines of massive particles must always
remain time-like, which implies that
\begin{equation}
1 / (1 + z) > 0
\end{equation}
on all time-like world lines over all regions of spacetime.
Therefore the matter equations of motion
determined by the Bianchi Identity $T^{\mu\nu};\nu = 0$,
where $T^{\mu\nu}$ is the energy-momentum tensor,
must always describe time-like motion.
This acts as a constraint on the radiation transport
described by the energy momentum tensor on the right hand sided of the
Einstein equation $G^{\mu\nu} = 8\pi T^{\mu\nu}/ c^4$.

It is clear that a strict application of the POE to the
solutions of the Einstein field equations in General
Relativity, implies that trapped surfaces leading to
event horizons cannot be physically realized through the time-like collapse of
radiating physical matter. If an event horizon were to form
during collapse, this would require the time-like world line
of the collapsing matter to become null in violation of the POE.
On the other hand, there is an enormous volume of literature
concerning black holes that is built entirely on the assumption that
trapped surfaces exist. In previous work
we have presented observational evidence against
this assumption. Here we provide a specific MECO model of
compact objects that is in accord with this POE requirement.

Although the existence of objects compact
enough to qualify as black hole candidates is beyond question,
no compelling evidence for the existence of
event horizons has been found. In fact,
It has recently been argued (Abramowicz, Kluzniak \& Lasota 2002),
that proof is fundamentally impossible, precisely because
highly red shifted, compact objects cannot be excluded. The reverse is not
true, however, for it may be possible to prove that these
compact objects possess intrinsic magnetic moments. 

Let us now consider the time-like collapse of a compact, radiating plasma
of mass $M_s$, surrounded by a physical surface, $R_s$ in contact with the
plasma. Then generalizing from earlier work, (Hernandez \&
Misner 1966; Lindquist, Schwartz \& Misner 1965; Lindquist 1966,
Misner 1965) the co-moving metric associated with this situation
will have the general form:
\begin{equation}
ds^2 = A^2c^2dt^2+2 DcdtdR-B^2 dR^2-R^2(t_r)(d\theta^2+ sin^2\theta d\phi^2)
\end{equation}
where $t_r = t - r/c$ is the retarded time.

Substitution of the metric into the Einstein equation
$G^{\mu\nu} = 8\pi T^{\mu\nu}/ c^4$,
with an energy momentum tensor $T^{\mu\nu}$ whose form
describes the radiating plasma collapse leads to a set of equations
from which, with appropriately chosen boundary conditions,
the functions A(r,t$_r$), B(r,t$_r$), and D(r,$t_r$) can be determined.
However, independently of the detailed form of the $T^{\mu\nu}$,
the first integral of the mixed time-time component of the Einstein
equation gives the proper time of the
collapsing surface synchronized along world lines of the
collapsing plasma on the surface, $R_s$ as
\begin{equation}
d\tau_s  = dt_r / (1 + z_s) =  dt_r (\Gamma_s + U_s/c) > 0
\end{equation}
where $z_s$ is the red shift of the radiating collapsing
surface, $t_r = t - R_s/c$  and
\begin{equation}
\Gamma_s = (1-\frac{R_{schw}}{R_s}+(\frac{U_s}{c})^2)^{1/2}.
\end{equation}
Here $U_s = (\frac{dR}{d\tau})_s < 0$  is the proper
time rate of change of the radius associated with the invariant circumference
of the collapsing surace, $R_s$ and $R_{schw} = 2GM_s/c^2$,
where $M_s(r ,t)$ is the total mass contained within the surface as
defined by the general formula:
\begin{equation}
M_s(R_s , t) = \int( T_0^0 \Gamma dV )_s.
\end{equation}
Here $\Gamma_s = (\frac{dR}{dl})_s$ and $dV$ is the proper volume element,
$dV = 4 \pi R^2 dl$. Here $dl$ is the proper length inside and on the
surface $R = R_s$ and is given by $dl = B dR$.

Since $M_s$ is being radiated away by photon emission then the
luminosity at infinity is given by
\begin{equation}
L_\infty =  -c^2\frac{dM_s}{dt_r}= -c^2\frac{dM_s}{d\tau}(1 + z_s)
\end{equation}
and the proper time rate of change of the Schwarzschild radius
with the mass $M_s$ inside is
\begin{equation}
U_{schw} = (\frac{dR_{schw}}{d\tau})_s  = \frac{2G}{c^2} \frac{dM_s}{d\tau_s} <  0
\end{equation}
Now since the POE requires that the collapse of the radiating physical
surface must always be time-like, it follows that the element of proper time
synchronized along the radial world line of a fluid element on the boundary
of the collapsing fluid must always obey $d\tau_s > 0$.
Hence this requires that (see Appendix A)
\begin{equation}
d\tau_s = dt_r((1-\frac{R_{schw}}{R_s}+(\frac{U_s}{c})^2)^{1/2}+\frac{U_s}{c})
    = \frac{dt_r}{1+z_s} > 0,
\end{equation}
Hence from Equation (10) we see that for $U_s <0$, the POE
requires that the \textit{`no trapped surface condition'}, $R_{schw}/R_s < 1 $ must hold.

For a collapsing plasma emitting particles at the local
Eddington limit, $L_{edd}$, the local radial collapse velocity of the
surface, $U_s$ essentially vanishes. Under these conditions the POE requirement
$R_s > R_{schw}$  for the collapse process, when differentiated
with respect to proper time, implies that
$0 = -U_s < -U_{schw}$ is dynamically satisfied by the
equations of motion of the matter.
It follows that the POE requires that the physical
dynamics associated with the general relativistic radiation
transfer process must prevent the collapsing surface from
passing through the Schwarzschild radius.
In Appendix A we show in detail that when the Eddington limit is
established at red shift $(1+z) = (1+ z_{edd})$, the POE
applied to the Einstein Equation implies that the time-like
collapsing radiating surface of the MECO lies outside
of the Schwarzschild radius of the collapsing object and remains that way
for the duration of the Eddington limited collapse process.
In Appendix B we show that if $(L(escape)_{edd})_s$ is the
luminosity generated within which escapes from the collapsing
surface, then the red shift at which an Eddington balance
will be achieved is given by
\begin{equation}
1 + z_{ Edd}  = \frac{\kappa(L(escape)_{edd})_s}{4 \pi G M_s c}.
\end{equation}
Here G is the Newtonian gravitational force
constant and $\kappa$ the plasma opacity.
Then in terms of the quantity $U_{schw}= \frac{dR_{schw}}{d\tau}$,
this implies that
\begin{equation}
\frac{U_{schw}}{c}=\frac{-2G (L(escape)_{edd})_s}{c^5 (1 + z_{edd})} < \frac{U_s}{c}
\end{equation}
consistent with the POE requirement that $R_{schw} < R_s$  must be maintained
by the physical forces which act during the collapse process.

If one naively attributes the Eddington limit luminosity to purely thermal
processes, one quickly finds that the required MECO surface temperatures
would be so high that photon energies would be well above the pair production
threshhold and extensive electron-positron pair production would
occur due to photon-photon collisions.
Thus the MECO surface region must be dominated by a pair plasma, and
at a temperature not greatly different from the $6\times 10^9 K$ threshhold
for photon-photon collisions to produce pairs (Pelletier \& Marcowith 1998).
In addition, a substantial
magnetic field is necessary in order to confine the pair plasma. For
a source of radius R and radiating luminosity L, a balance of radiative and
magnetic stresses would require a magnetic field
strength of $B > \sqrt{4L/\pi R^2c} \sim 10^8$ gauss for
a 10 M$_\odot$ Eddington limit GBHC. For an Eddington limit L increasing as 1+z,
as implied by equation 11, the surface magnetic field would need to be larger by
the factor of $\sqrt(1+z)$. For the large red shifts contemplated in this work,
it will be necessary to consider very large surface magnetic fields.
Pelletier \& Marcowith (1998) have shown that the energy of magnetic
perturbations in pair plasmas can be preferentially transferred to radiation,
rather than causing particle acceleration for equipartition magnetic fields.
The radiative power of
an equipartition pair plasma is proportional to $B^4$ because
the pair density is proportional to $B^2$,
in addition to a synchrotron energy production proportional to $B^2$.
Because of this extremely efficient photon production mechanism, the radiation
temperature is buffered. When particle kinetic pressure is in equipartition
with magnetic pressure, the field also exceeds that required to confine the
plasma, thus stability is maintained.

In an extremely compact object, the particle kinetic pressure might be
as high as $\rho c^2/3$. If we take $\rho = M/(4\pi R^3/3), R\approx 2GM/c^2)$
and equate the magnetic and particle kinetic pressures, we obtain
$B_{equip} = c^4/(4MG^{3/2}) = 6\times 10^{18}/m$ gauss,
where $m=M/M_\odot$ is the mass in solar mass units.
Thus the dynamic condition required for an Eddington limited stationary
equilibrium to initially occur at red shift $(1 + z_{edd})$ during the
MECO plasma collapse process is that the MECO must contain an
equipartition magnetic field with energy density $B^2/8\pi$ that is
comparable to the particle kinetic pressure. Then
the existence and stability of the $U_s = 0$,
Eddington limited MECO regime is guaranteed because the intrinsic,
equipartition magnetic field is the primary source of luminosity. Since
this luminosity is not confined to the core of the MECO it will not
be trapped, as occurs with neutrinos, however, the radiation should be
thermalized by the optically thick environment from which it escapes.
Finally, we note that although there have been many numerical computations
that apparently show the formation of trapped surfaces, to our knowledge
none have had sufficient numerical resolution to examine this extreme
red shift regime nor have they considered the emergent properties of
equipartition magnetic fields and pair plasmas at high red shift.
As we shall see, an electron-positron pair atmosphere of
a MECO is an extremely significant structure that
conveys radiation from the MECO surface to a zone with a much lower
red shift and larger escape cone from which it escapes. In order to capture it within a
numerical grid, a grid point spacing of at least $10^{-8}R_g$ would be needed,
where $R_g = GM/c^2$ is the gravitational radius.

The strength of the intrinsically generated magnetic fields $B_{env}$
observed in the distant environment around the MECO are reduced
by a factor of $(1+z_{edd})$
from their values near $R_s$. The fields needed to produce jets
in AGN are observed to be of the order $10^3 - 10^4$ gauss as judged from
a distance. On the other hand, a distantly observed equipartition field would
be $\sim 6\times 10^{18}/(m(1+z_{edd}))$ gauss. This suggests that for an
$m \sim 10^8$ AGN, the combined effect of mass scaling and red shift would need
to reduce the surface field from $6\times 10^{18}$ gauss to $10^{3 - 4}$ gauss. This
would require the MECO to have a red shift of $z \sim 10^7 - 10^8$.
In previous work, (Robertson \& Leiter 2002) we have found typical
magnetic fields of a few times $10^{10}$ gauss for GBHC. These would
require similar values of $z_{edd} \sim 10^8$, as well as $ m\sim 10$.
Therefore for both GBHC and AGN we find that
\begin{equation}
1 + z_{edd} = \frac{B_{equip}}{B_{env}} \sim  10^8.
\end{equation}
\footnote{An additional point of support for very large values of z
concerns neutrino transport in stellar core collapse. If a diffusion limited
neutrino luminosity of $\sim 10^{52}$
erg/s (Shapiro \& Teukolsky 1983) were capable
of briefly arresting the collapse, then the subsequent reduction of
neutrino luminosity as neutrino emissions are depleted in the core would lead to
a rapid adiabatic collapse until photon emissions reach an Eddington limit.
At this point the photon luminosity
would need to support a smaller and much more tightly gravitationally
bound mass. An order of magnitude calculation of binding energy
($\sim Mc^2 ln(1+z)$) indicates that when stable, an m = 10 MECO
should be approximately 17 times less massive than at the point of loss
of neutrino support. (This has obvious, important consequences for
hypernova models of gamma ray bursters.) A new photon Eddington
balance would thus require an escaping luminosity reduced by
a factor of 17 and also reduced by the ratio $(\sigma_T/\sigma_\nu)$,
where $\sigma_T = 6.6\times 10^{-25}$ cm$^2$ is the Thompson cross section and
$\sigma_\nu = 4.4\times 10^{-45}$ cm$^2$ is the neutrino scattering
cross-section. (This would be the opacity ratio as long as particles
of the same mass are being supported by both photons and neutrinos.)
Thus $L_\infty \sim 8\times 10^{32}$ erg/s would be
required. For a 7 M$_\odot$ GBHC, this would require $1+z \sim 10^8$
It is of some interest that neutrinos with non-zero rest mass would
be trapped inside the photon sphere anyway.}

The quiescent luminosity of a MECO originates deep within its photon sphere.
When distantly observed it is diminished by both gravitational red shift
and a narrow exit cone. The gravitational red shift
reduces the surface luminosity by $1/(1+z)^2$ while the exit cone further
reduces the luminosity by the factor
$27(1+z)^2 u^2 \sim 27(1+z)^2/4$ for large z. (See Appendix C).
Here we have used
\begin{equation}
u = \frac{GM_s}{r c^2} = \frac{R_g}{r} = \frac{1}{2}(1 - \frac{1}{(1+z)^2})
\end{equation}
where r and z refer to the location from which photons escape.
The net outflow fraction of the luminosity  provides the support for
the collapsing matter. The photons actually escape from the photosphere
of a pair atmosphere.

\section{The quiescent MECO}
The fraction of luminosity from the MECO surface
that escapes to infinty in Eddington balance is (Appendix B):
\begin{equation}
(L_{edd})_s = \frac {4\pi G M_s c(1+z)}{\kappa} = 1.27 \times 10^{38}m(1+z_s)~~~~ erg/s
\end{equation}
In the last expression we have used $\kappa = 0.4$ cm$^2$/g.
The distantly observed luminosity is:
\begin{equation}
L_\infty = \frac{(L_{edd})_s}{(1+z_s)^2} = \frac {4\pi G M_s c}{\kappa(1+z_s)}
\end{equation}
When radiation reaches the photosphere, where the temperature is $T_p$,
the fraction that escapes is:
\begin{equation}
L_\infty = \frac{4 \pi R_g^2 \sigma T_p^4}{u_p^2} \frac{27 u_p^2}{(1+z_p)^4}
    = 1.56\times 10^7 m^2 T_p^4 \frac{27}{(1+z_p)^4}~~~~erg/s
\end{equation}
where $\sigma = 5.67\times 10^{-5}$ erg/s/cm$^2$
and subscript p refers to conditions at the photosphere.
Equations 16 and 17 yield:
\begin{equation}
T_\infty = T_p/(1+z) = \frac{2.3\times 10^7}{(m(1+z_s))^{1/4}}~~~~ K.
\end{equation}
To examine typical cases, a $10 M_\odot$, $m = 10$ GBHC modeled in
terms of a MECO with $z \sim 10^8$ would have
$T_\infty = 1.3\times 10^5 K = 0.01$ keV, a luminosity of
$L_\infty =1.3\times 10^{31} erg/s$, and a spectral peak at 220 A$^0$,
in the photoelectrically absorbed deep UV.
For an m=$10^7$ AGN, $T_\infty = 4160 K$, $L_\infty = 1.3\times 10^{37} erg/s$
and a spectral peak in the near infrared at 7000 A$^o$.
Considering these emission rates as
indicators of the rate that high red shift MECO would lose mass, their
apparent radiative lifetimes would be millions of Hubble times.
(See Appendix A \& B for exact lifetime results).
Hence passive MECO without active accretion disks, although
not black holes, have lifetimes much greater
than a Hubble time and emit highly red
shifted quiescent thermal spectra that would be quite
difficult to observe. There are additional power law components
of similar magnitude that originate as magnetic dipole
spin-down radiation (see below).

Escaping radiation passes through a pair plasma atmosphere that can be shown,
\textit{ex post facto} (See Appendix F), to be radiation dominated throughout.
Under these circumstances, the radiation pressure within the equilibrium
atmosphere obeys $P_{rad}/(1+z) = constant$. \footnote{We
consider the pair atmosphere to exist external to the Meco. In
exterior Schwarzschild geometry, the hydrostatic balance
equation within the MECO atmosphere is
$\frac{\partial p}{\partial r} =
-\frac{\partial \ln{(g_{00})}}{2 \partial r}(p + \rho c^2)$, where
$g_{00} = (1-2u)$ and $\rho c^2 << p$. This integrates to $p/(1+z) = constant$.}
Thus the relation between surface and photosphere temperatures is
$T_s^4/(1+z_s) = T_p^4/(1+z_p)$. At the MECO surface, we expect a
pair plasma temperature of $T_s \approx m_ec^2/k \sim 6\times 10^9 K$ because
an equipartition magnetic field effectively acts as a thermostat which
buffers the temperature of the optically thick synchrotron radiation
escaping from the MECO surface (Pelletier \& Marcowith 1998).
But since $T_\infty = T_p/(1+z_p)$,
we have that
\begin{equation}
T_p = T_s(\frac{T_s}{T_\infty (1+z_s)})^{1/3}
= \frac{1.76\times 10^9}{(m(1+z_s))^{1/12}} ~~~~K
\end{equation}
For $1+z_s=10^8$ and $m=10$ GBHC, this yields a photosphere temperature of
$3.1\times 10^8$ K, from which $(1+z_p) = 2400$. An AGN with $m=10^7$
would have a somewhat cooler photosphere at $T_p = 9.9\times 10^7$ K, but
with a red shift of 24000. If surface temperature for an AGN MECO
were somewhat lower than $6\times 10^9 K$, the pair mass density
could be below the mean density of matter in the AGN. Thus it becomes
plausible to consider that AGN might be predominately pair plasma with
relatively small baryonic content.

\section{An Actively Accreting MECO}
From the viewpoint of a distant observer, accretion would
deliver mass-energy to the MECO, which would then radiate most of it
away. The contribution from the central MECO alone would be
\begin{equation}
L_\infty = \frac {4\pi G M_s c}{\kappa(1+z_s)}+ \frac{\dot{m}_\infty c^2}{1+z_s}(e(1+z_s)-1)
    = 4 \pi R_g^2 \sigma T_p^4 \frac {27} {(1+z_p)^4}
\end{equation}
where $e = E/mc^2 = 0.943$ is the specific energy per particle
available after accretion disk flow to the marginally stable orbit radius,
$r_{ms}$. Assuming that $\dot{m}_\infty$ is some fraction, f, of the
Newtonian Eddington limit rate, $4\pi G M c/\kappa$, then
\begin{equation}
1.27\times 10^{38}\frac{m\eta}{1+z_s} =
(27)(1.56\times 10^7)m^2(\frac{T_p}{1+z_p})^4
\end{equation}
where $\eta=1+f((1+z_s)e -1)$ includes both quiescent and accretion
contributions to the luminosity.
Due to the extremely strong dependence on temperature of the
density of pairs, (see Appendix F) it is unlikely that the temperature
of the photosphere will be greatly different from the $3.1\times 10^8
K$ found previously for a typical GBHC. Assuming this to be the case, along with
$z=10^8$, $m=10$, and $f=1$, we find $T_\infty = T_p/(1+z_p) = 1.3\times 10^7 K$
and $(1+z) = 24$, which indicates considerable photospheric expansion.
The MECO luminosity would be approximately
$L_\infty = 1.2\times 10^{39}$ erg/s. For comparison, the accretion disk
outside $r_{ms}$ (efficiency = 0.057) would produce only $6.8\times 10^{37}$ erg/s.
Thus the high accretion state luminosity of a GBHC would originate
primarily from the central MECO. A substantial fraction of the
softer thermal luminosity would be Compton scattered to higher energy
in the plunging flow inside $r_{ms}$. The thermal component
would be `ultrasoft' with a temperature of only $1.3\times 10^7 K$.
Even if a disk flow could be maintained all the way to the MECO
surface, where a hot equatorial band might result, the escaping
radiation would be spread over the larger area of the photosphere due
to photons origins deep inside the photon orbit.

For radiation passing through the photosphere
most photons would depart with some azimuthal momentum on spiral trajectories
that would eventually take them across and through the accretion disk.
Thus a very large fraction of the soft photons would be subject
to bulk comptonization in the inflow from the accretion disk.
This contrasts sharply with the situation for neutron stars
where few photons from the surface cross the disk. This
could account for the fact that hard x-ray spectral tails are
comparatively much stronger for high state GBHC.
Our preliminary calculations for photon trajectories
randomly directed upon leaving the
photon sphere indicate that this process
would produce a power law component with photon index greater than 2.
These are difficult and important calculations for which the effects
of multiple scattering are crucial. But they are beyond the
scope of this paper, which is intended as a first description of the
general MECO model.

\section{Discussion}
Since the MECO model provides a framework for understanding most
of the known spectral and timing features of compact x-ray sources, it
is useful to recapitulate important features.
The progression of configurations of accretion disk, magnetic field
and boundary layer is shown in Figure 1. We begin with
quiescence.

Quiescent luminosities that are generally 10 - 100 X lower for GBHC
than for neutron stars (NS) have been claimed as evidence for the existence
of event horizons. (Narayan et al. 1997, Garcia et al. 2002).
In our MECO model, the quiescent emissions are magnetic dipole
emissions that are characteristic of the magnetic moment and rate of
spin of the central object. The lower quiescent luminosities of the
GBHC are explained by their lower spin rates and (perhaps unobservably) low
rates of quiescent emission from the central MECO.

In previous work (Robertson \& Leiter 2002) we found that
magnetic moments and spin rates could be determined from luminosities
at the end points of a spectral hardening transition. This spectral
state switch for NS in low mass x-ray binaries (LMXB) is due to a magnetic
propeller effect (Ilarianov \& Sunyaev 1975, Stella, White
\& Rosner 1986, Cui 1997, Zhang, Yu \&
Zhang 1997, Campana et al. 1998). The magnetic moments and spins
were used to calculate the soft x-ray luminosity expected from
low state spin-down. The results are recapitulated and extended
in Table 1. The equations and methods of calculation
are repeated, with minor corrections, in Appendix D. Calculated
values of quiescent luminosity in Table 1 have been corrected using
a more recent correlation of spin-down energy loss rate and soft
x-ray luminosity (Possenti et al. 2002), but results are otherwise unchanged
from the previous work except for new additions.
It is a very powerful confirmation of the propeller mechanism that spins are
in good agreement with burst oscillation frequencies (Strohmayer 2000),
magnetic moments are
of similar magnitude to those determined from the spin-down of
millisecond pulsars and the calculated quiescent luminosities are accurate.

Even though the quiescent surface luminosity of the MECO is very low,
surface and the magnetospheric spin-down luminosities are capable of ablating the
material in a quiescent accretion disk. For a GBHC, radiation at
$\sim 10^{31}$ erg/s should raise the temperature of the optically
thick inner disk above the $\sim 5000$ K instability temperature for
hydrogen out to a distance of $r \sim 10^{10}$ cm. Therefore we expect the
quiescent inner disk to be essentially empty with a large inner radius.
The rate of mass flow ablated at the inner disk radius would only
need to be $\sim 10^{13}$ g/s to produce the
quiescent optical emission observed for GBHC and NS. The ablated material
could escape if it reached the magnetic propeller region, which is
confined to the light cylinder at a much smaller radius than that of the
inner disk. This makes the MECO model
compatible with the disk instability model of x-ray nova
outbursts, which begin as `outside-in' events in which substantial outer mass
reservoirs have been observed to  fill an accretion disk on the
viscous timescale of a radially subsonic flow (Orosz et al. 1997).

In outburst, the disk flow first engages
the magnetic field of the rotating central object near the light cylinder
radius, $r_{lc} = c/2\pi \nu_s$. A boundary layer forms in the disk where
matter of the inner radius is, at least temporarily, brought into co-rotation
with the magnetosphere and loaded onto its field lines. Behind the disk
boundary layer, the flow remains Keplerian and largely shielded
by induced surface currents from the MECO magnetic field.
As the inflow proceeds, the magnetosphere rejects it via the
`propeller effect' until the inner disk
can push inside the co-rotation radius, $r_c$. From $r_{lc}$ to
$r_c$, the system is in the Low/Hard spectral state. Inside $r_c$, the
propeller regime ends and matter of sufficient pressure can make its way inward.
From quiescence to the light cylinder, the
x-ray luminosity changes by a factor of only a few as the disk
generates a soft thermal spectral component (which may be mistaken
for surface radiation for NS.) From $r_{lc}$ to $r_c$, the x-ray luminosity
may increase by a factor of $\sim 10^3 - 10^6$. With inner disk inside
$r_c$, the outflow and/or jets subside, the system becomes
radio quiet, the photon index increases, and a soft
thermal excess appears, both of which
contibute to a softer spectrum, (e.g., see Fig. 3.3 of Tanaka \& Lewin, p. 140),
which may be even be described as `ultrasoft' (White \& Marshall 1984);
particularly when the luminosity finally begins to decline.

Plasma flowing outward in the low state may depart in a jet, or as an outflow
back over the disk as plasma is accelerated on outwardly curved magnetic field
lines. Radio images of both flows have been seen (Paragi et al. 2002).
Equatorial outflows could contribute to the low state hard spectrum
by bulk Comptonization of soft photons in the outflow. This would
accentuate the hardness by the depletion of the soft photons that would
otherwise be observed to arise from the disk. Such an outflow
would be compatible with partial covering models for dipping sources, in which
the hard spectral region seems to be extended and of small extent
perpendicular to the disk (Church 2001, Church \& Balucinska-Church 2001).
Alternatively, an accretion disk corona might be a major contributor to
the hard spectrum. For jet emissions, recent work (Corbel
\& Fender 2002) has shown that it may be possible to explain much of the
broadband emissions from near infrared through soft x-rays as
the power-law synchrotron emissions of compact jets, which
have been directly imaged for some GBHC. Jets would be compatible with
`lamp post' reverberation models of AGN. It is noteworthy that strong
($> 10^8$) gauss magnetic fields have been found to be necessary at the
base of the jets of GRS 1915+105 (Gliozzi, Bodo \& Ghisellini 1999,
Vadawale, Rao \& Chakrabarti 2001). A recent study of optical polarization
of Cygnus X-1 in its low state (Gnedin, Silantev \& Titarchuk 2002)
has found a magnetic field of $\sim 10^7$ gauss at the location of the optical
emission. These fields at distances approximately equal to the
co-rotation radius imply magnetic moments for both GRS 1915+105 and Cygnus X-1
that are in good agreement with those of Table 1.
Either jet or equatorial outflows would appear to be manifestations of
the interaction of the magnetic field of the central object and the accretion
disk. 

\begin{table*}
\begin{center}
\caption{$^a$Calculated and Observed Quiescent Luminosities}
\begin{tabular}{lrrrrrrrr} \hline
Object & m & $L_{min}$ & $L_c$ & $\mu_{27}$ & $\nu_{obs}$ & $\nu_{calc}$ & log (L$_q$) & log (L$_q$) \\
    & M$_\odot$ & $10^{36}$erg/s & $10^{36}$erg/s & Gauss cm$^3$    & Hz      &  Hz~ & erg/s & erg/s \\
    & & & & & obs. & calc. & obs. & calc. \\ \hline
\bf {NS} \\
Aql X-1 & 1.4 & 1.2 & 0.4 & 0.47 & 549 & 658 & 32.6  & 32.5  \\
4U 1608-52& 1.4 & 10 &2.9 & 1.0 & 619 & 534 & 33.3 & 33.4 \\
Sax J1808.4-3658& 1.4 & $^b$0.8 & 0.2 & 0.53 & 401 & 426 & 31.8-32.2 & 32 \\
Cen X-4 & 1.4 & 4.4 & 1.1 & 1.1 & & 430 & 32.4 & 32.8 \\
KS 1731-26 & 1.4 & & 1.8 & 1.0 & 524 & & $^c$\bf{32.8} & 33.1  \\ 
\bf{XTE J1751-305} & 1.4 & & 3.5 & 1.9 & 435 & & $<$34.3 & 33.7 \\
\bf{XTE J0929-314} & 1.4 & & 4.9 & 8.5 & 185 & & & 33.1 \\
4U 1916-053 & 1.4 & $\sim$14 & 3.2 & 3.7 & 270 & 370 & & 33.0 \\
\bf{4U1705-44} & 1.4 & 26 & 7 & 2.5 & & 470 & & 33.7 \\
4U 1730-335 & 1.4 & 10 & & 2.5 & 307 & & & 32.9 \\
\bf{GRO J1744-28} & 1.4 & & 18 & 13000 & 2.14 & & & 31.5 \\
Cir X-1 & 1.4 & 300 & 14 & 170 & & 35 &  & 32.8 \\ \hline
\bf{GBHC} \\
GRS 1124-68 & 5 & 240 & 6.6 & 720 & & 16 & $<32.4$ & 32.7 \\
GS 2023+338 &7 & 1000 & 48 & 470 & & 46 & 33.7 & 34 \\
XTE J1550-564 & 7 & $^d$90 & 4.1 & 150 & & 45 & 32.8 & 32.2 \\
GS 2000+25 & 7 & & 0.15 & 160 & & 14 & 30.4 & 30.5\\
GRO J1655-40 & 7 & 31 & 1.0 & 250 & & 19 & 31.3 & 31.7 \\
A0620-00 & 4.9 & 4.5 & 0.14 & 50 &  & 26 & 30.5 & 30.2 \\
Cygnus X-1 & 10 & & 30 & 1260 & & 23 &  & 33 \\
GRS 1915+105 & 7 & & 12 & 130 & $^e$67 & & & 33 \\
\bf{XTE J1118+480} & 7 & & 1.2 & 1000 & & 8 & & 31.5 \\
\bf{LMC X-3} & 7 & 600 & 7 & 860 & & 16 & & 33 \\ \hline
\end{tabular}
\end{center}
$a$New table entries in bold font are described in Appendix E. \\
Equations used for calculations of spins, magnetic moments and $L_q$ are in Appendix D.\\
Other tabular entries and supporting data are in Robertson \& Leiter (2002)\\
$^b$2.5 kpc, $^c$(Burderi et al. 2002), $^d$d = 4 kpc \\
$^e$GRS 1915+105 Q $\approx 20$ QPO was stable
for six months and a factor of five luminosity change. \hfill \\
\end{table*}

The Intermediate/Very High State occurs as the inner disk region moves
inside $r_c$. This passage is often accompanied by substantial mass
ejection in jets. It is followed by the Soft/High state in which accreting
matter can flow to the central object. For matter suffiently
inside $r_c$, the propeller mechanism is incapable of stopping the flow,
however, a boundary layer may form at the inner disk radius in this case.
The need for a boundary layer for GBHC can be seen by comparing the magnetic
pressure at the magnetosphere with the impact pressure of a trailing,
subsonic disk. For example, for an average GBHC magnetic moment of
$\sim 4\times10^{29}$ gauss cm$^3$ from Table 1, the magnetic pressure
at a $r_{ms}$ radius of $6.3\times10^6$ cm for a 7 M$_\odot$
GBHC would be $B^2 /8\pi \sim 10^{17}$ erg/cm$^3$. At a mass flow rate of
$\dot{m} = 10^{18}$ g/s, which would be near Eddington limit
conditions for a 7 M$_\odot$ MECO, the inner disk temperature would be
$T \sim 1.5\times10^7$ K. The disk scale height would be given by
$H\sim r v_s/v_K \sim 0.0036r$, where $v_s \sim 4.5\times 10^7$cm/s
and $v_K \sim 1.2\times 10^{10}$ cm/s are acoustic and
Keplerian speeds, respectively. The impact pressure
would be $\dot{m}v_r/4\pi r H \sim 5.6\times10^5 v_r$ erg/cm$^3$.
It would require $v_r$ in excess of the speed of light to
let the impact pressure match the magnetic pressure.
But since the magnetic field doesn't move fast enough to eject the
disk material inside $r_c$, matter piles up as essentially dead weight against
the magnetopause and pushes it in. The radial extent of such a layer
would only need to be $\sim kT/m_pg \sim 50$ cm,
where $m_p$ is the proton mass and
$g$, the radial gravitational free fall acceleration, but it is likely
distributed over a larger transition zone from co-rotation with the magnetosphere
to Keplerian flow. The gas pressure in the inner part of the
transition zone necessarily matches the magnetic
pressure. We observe that if this be the case, radiation pressure in the disk,
at $T= 1.5\times 10^7 K$,
is nearly three orders of magnitude below the gas pressure. Therefore a
gas pressure dominated, thin, Keplerian disk with subsonic radial speed
should continue all the way to
$r_{ms}$ for a MECO. Similar conditions occur with disk radius inside $r_c$
even for weakly magnetic NS.
The nature of mass accumulations in the inner disk
transition region and the way that they can enter the magnetosphere
have been the subject of many studies, (e.g., Spruit \& Taam 1990).

In the case of NS, sufficiently high mass accretion rates
can push the magnetopause into the star surface. At this point the
hard apex of the right side of the horizontal branch of the `Z'
track in the hardness/luminosity diagram is reached. It has recently been
shown (Muno et al. 2002) that the distinction between `atoll' and
`Z' sources is merely that this point is reached near the Eddington
limit for `Zs' and at perhaps $\sim 10 - 20$\% of this luminosity
(Barrett \& Olive 2002) for the less strongly magnetized `atolls'.
Atolls rarely reach such luminosities. For MECO based GBHC, one
would expect a relatively constant ratio of hard and soft x-ray
`colors' (e.g. van der Klis, et al.) after the inner disk crosses $r_c$ and the
flow reaches the photon orbit.

For the more massive GBHC and AGN, the disk, when inside $r_c$, is not masked
by outflow and the disk itself shows a soft thermal spectral component.
However, as we have shown in Section 4, significantly brighter
radiation of similar temperature arises from matter
plunging inside $r_{ms}$ and reaching the MECO.

An observer at coordinate, r, inside $r_{ms}$, would find the
radial infall speed to be $v_r = \frac{\sqrt{2}}{4}c(6u-1)^{3/2}$,
where $u = GM/rc^2 = R_g/r$ (see Appendix C)
and the Lorentz factor for a particle spiraling
in from $6R_g$ would be $\gamma = 4 \sqrt{2}(1 + z)/3$, where
$1+ z = (1-2u)^{-1/2}$ would be the red shift for photons generated at $r$.
If the distantly observed mass accretion rate would be $\dot{m}_\infty$, then
the impact pressure at r would be $p_i = (1+z)\dot{m}_\infty \gamma v_r/ (4\pi r H)$.
For $\dot{m}_\infty \sim 10^{18}$ g/s, corresponding to Eddington limit
conditions for a 7 M$_\odot$ GBHC, and $H = 0.0036r$, impact pressure is,
$p_i \sim 5 \times 10^{16}(1 + z)^2(2R_g/r)^2(6R_g/r-1)^{3/2}$
erg/cm$^3$. For comparison, the magnetic pressure
is $(1 + z)^2 B_\infty^2/8\pi$. Assuming a dipole
field with average magnetic moment of $4\times 10^{29}$ gauss cm$^3$
from Table 1, the magnetic
pressure is $\sim 10^{20} (1+z)^2 (2R_g/r)^6$ erg/cm$^3$.
There are no circumstances for
which the impact pressure is as large as the
magnetic pressure for $2R_g < r < 6R_g$.
Thus we conclude that another weighty boundary layer must form inside
$r_{ms}$ in order to push the magnetosphere inward.
The inner radius of the disk is determined by the rate at which
the magnetic field can strip matter and angular momentum from the disk.
This occurs in a boundary layer of some thickness, $\delta r$, that
is only a few times the disk thickness. (See Appendix D)

Other than the presence of a transition boundary layer on the magnetopause,
the nature of the flow and spectral formation inside $r_c$ is a research topic.
Both the short distance from $r_c$ to $r_{ms}$ and the
magnetopause topology should help to maintain a
disk-like flow to $r_{ms}$. Radial acceleration inside $r_{ms}$
should also help to maintain a thin structure. These flows are depicted
in Figure 1. As discussed in Section 4, we expect the flow into the
MECO to produce a distantly observed soft thermal component, part of which is
strongly bulk Comptonized.

Although many mechanisms have been proposed for the high frequency
quasi-periodic oscillations (QPO) of x-ray luminosity, they
often require conditions that are incompatible with thin, viscous
Keplerian disks. Several models have requirements for
lumpy flows, elliptical inner disk boundaries,
orbits out of the disk plane or conditions
that should produce little radiated power. In a conventional thin disk,
the vertical oscillation frequency, which is approximately the same
as the Keplerian frequency of the inner viscous disk radius should
generate ample power.  Accreting plasma should periodically wind the poloidal
MECO magnetic field into toroidal configurations until the field lines break
and reconnect across the disk. Field reconnection across the
disk should produce high frequency oscillations that couple to the
vertical oscillations. There would be an automatic association of high
frequency QPO with magnetospherically driven power law emissions, as is
observed. Mass ejection in low state jets might be related to
the heating of plasma via the field breakage mechanism, in addition
to natural buoyancy of a plasma magnetic torus in a poloidal external field.

It seems possible that toroidal winding of field lines at the magnetopause,
breakage and reconnection might continue in high states inside $r_{ms}$.
If so, there might be QPO that could be identified as signatures of
the MECO magnetosphere. If they occur deep within the magnetosphere, they
might be at locally very high frequencies, and be observed distantly
red shifted as very low frequencies. As shown in Appendix C, the
`Keplerian' frequencies in the plunging region inside $r_{ms}$ are
given by $\nu = 1.18\times 10^5 u^2(1-2u)/m$ Hz. A maximum frequency of
437 Hz would occur for m=10 at the photon orbit. Of more interest, however
are frequencies for $u \approx 1/2$, for which $\nu = 2950/(m(1+z)^2)$ Hz.
For $1+z = 10, m=10$; conditions that might apply to the photosphere region,
$\nu \sim 3$ Hz could be produced.

Even if QPO are not produced inside $r_{ms}$ or inside the photon sphere
for GBHC, there is an interesting scale mismatch that might allow them
to occur for AGN. Although the magnetic moments of AGN scale inversely
with mass, the velocity of plasma inside $r_{ms}$ does not. Thus the
energy density of disk plasma inside $r_{ms}$ will be relatively larger than
magnetic field energy densities for AGN accretion disks. When field
energy density is larger than kinetic energy density of matter, the field
pushes matter around. When the reverse is true, the matter drags the field
along. Thus toroidal winding of the field at the magnetopause
could fail to occur for GBHC, but might easily do so for AGN.
If the process is related to mass ejection, then very energetic jets
with Lorentz factors $\gamma \sim (1+z)$ might arise from within
$r_{ms}$ for AGN. A field line breakage model of `smoke ring'
like mass ejection from deep within $r_{ms}$ has been developed by
Chou \& Tajima (1999). In their calculations, a pressure of unspecified
origin was needed to stop the flow outside $r_{schw}$ and a poloidal
magnetic field, also of unspecified origin was required. MECO
provide the necessary ingredients in the form of the intrinsic
MECO magnetic field. The Chou \& Tajima mechanism is apparently
not active inside $r_{ms}$ for GBHC, as their jet emissions appear to be
associated with the low/hard state (Pooley \& Fender 2002).

Finally, some of the rich oscillatory behavior of GRS 1915+105 may be
readily explained by the interaction of the inner disk and the central
MECO. The objects in Table 1 have co-rotation radii of order $10 R_{schw}$,
which brings the low state inner disk radius
in close to the central object. A low state MECO, balanced near
co-rotation would need only a small increase of mass flow rate to
permit mass to flow on to the central MECO. This would produce more than
20X additional luminosity and enough radiation pressure to blow the
inner disk back beyond $r_c$ and load its mass onto the magnetic field
lines where it is ejected. This also explains the association of jet
outflows with the oscillatory states. Belloni et al. (1997) have shown
that after ejection of the inner disk, it
then refills on a viscous time scale until the process repeats.
Thus one of the most enigmatic GBHC might be understood as a relaxation
oscillator, for which the frequency is set by a critical
mass accretion rate.

\section{Conclusions}
We have shown that the geodesics of physical matter would
become null, in violation of the POE, if trapped surfaces of infinite red
shift exist (e.g., Equations 1 - 3 and Appendix C).
An enormous body of physics scholarship developed
primarily over the last half century has been built on the assumption that
trapped surfaces exist. Misner, Thorne \& Wheeler (1973),
for example in Sec. 34.6 clearly state that this is an assumption and that
it underlies the well-known singularity theorems of Hawking
and Penrose. In contrast,
we have asserted the \textit{ `no trapped surface condition'} and found new,
quasi-stable, high red shift MECO solutions of the Einstein field equations.
The physical mechanism of stability is an Eddington balance maintained
by the distributed photon generation of an equipartition magnetic field.
This field also serves to confine the pair plasma of the outer layers of
the MECO and the MECO pair atmosphere. Red shifts of $z \sim 10^8$
have been found to be necessary for compatibility with our previously
found magnetic moments for GBHC.

Strong magnetic fields are the robust hallmarks of MECO.
Their existence in GBHC and AGN is implied by their synchrotron radiations,
and is also clearly shown by the correlated data in Table 1. The spectral state
switch is the signature of the magnetic field in its interaction
with an accretion disk. Our work, as well as that of others, has shown
that the magnetic propeller effect is the mechanism of the
spectral state switch. These changes of spectral characteristics
and luminosity have been observed for AGN, GBHC and NS,
for which there are independent, confirming measurements of spin that
show the switch to be taking place with the inner radius of the accretion
disk at the co-rotation location.
For the NS, the agreement between calculated and observed
quiescent luminosities confirms that the quiescent power law emissions are
magnetic dipole radiation. Since magnetic dipole radiation is proportional to
the fourth power of the spin, the lower quiescent luminosities of the GBHC
are simply a result of their slower spins.

It may be possible to observe MECO in several other ways. Firstly, as we
have shown, for a red shift
of $z \sim 10^8$, the quiescent luminosity of a GBHC MECO would be
$\sim 10^{31}$ erg/s with $T_\infty \sim 0.01$ keV. This thermal peak might be
observable for nearby  or high galactic latitude GBHC,
such as A0620-00 or XTE J1118+480. 
Secondly, at moderate luminosities $L \sim 10^{36} - 10^{37}$ erg/s but
in a high state at least slightly above $L_c$,
a central MECO would be a bright, small central object that might
be sharply eclipsed in deep dipping sources. A MECO should stand out
as a bright point source. A conclusive demonstration that the most of the soft
luminosity of a high state GBHC is distributed over a large radius would be
inconsistent with MECO or any other GBHC model entailing a central
bright source. Thirdly, a pair plasma atmosphere in an
equipartition magnetic field should be virtually transparent to photon
polarizations perpendicular to the magnetic field lines. The x-rays
from the central MECO should exhibit some polarization that might be
detectable, though this is far from certain since the distantly observed
emissions could originate from nearly any point on the photosphere.
Fourthly, an equipartition magnetic moment in a slowly rotating MECO
might not necessarily be aligned parallel to the spin axis. It might
be possible to observe pulsar oscillations under some circumstances. This
possibility should be tempered with the observation that most NS in LMXB
exhibit no pulsations in either x-ray or radio bands
despite their magnetic moments. Finally, MECO presumably would not
be found only in binary systems. If they are the offspring of massive
star supernovae, they should be found all over the galaxy. If we
have correctly estimated their quiescent temperatures, isolated MECO
would be weak, possibly polarized, EUV sources.

The MECO model, based on the firm ground of the Principle of Equivalence,
represents a paradigm shift for astrophysics. As a unified model, it
provides a natural explanation of the `ultrasoft'
high state thermal spectrum. The high state power law emissions
are due to bulk comptonization of photons from the MECO photosphere
as their spiral trajectories take most of them across the flow inside $r_{ms}$.
The weaker power law emissions of NS are a result of less source compactness and
fewer photons from the surface having trajectories that cross the disk.
The MECO model includes a magnetic moment and a
mechanism for the spectral state switch. It accounts for low state jets
and equatorial outflows as magnetic propeller effects. Its empty
inner disk in quiescence is consistent with the disk instability
model with its viscous timescale for the interval between optical and
x-ray brightening during outbursts. The accretion disk is dominated
by gas pressure in all states. The MECO model accounts for the power law
portions of the quiescent luminosity as magnetospheric spin-down
emissions. As we have shown in Table 1, the spectral state switch,
spin, magnetic moment and quiescent luminosity are firmly linked.
The low state jets and outflows, with their synchrotron
emissions, are obviously linked to strong magnetic fields.
It is generally accepted that an intrinsic magnetic moment
in the central object is completely consistent with the behaviour of NS.
The spectral, timing and synchrotron emission similarities of NS and GBHC
are also well known (e.g. Tanaka \& Shibazaki 1996, van der Klis 1994).
It strains credulity to think that GBHC can duplicate these richly
complex and obviously magnetic phenomena with just an event
horizon and a tricky disk.

\acknowledgements

\clearpage

\appendix{\bf{A. Eddington Limited Collapse for MECO}}\\
Let us assume an energy momentum tensor which involves matter, pressure, and
radiation
\begin{equation}
T_\mu^\nu = (\rho+ P/c^2) u_\mu u^\nu -P\delta_\mu^nu + E_\mu^\nu
\end{equation}
where $u^\mu = (u^0 , 0)$ in the interior where
co-moving coordinates are used and
$E_\mu^\nu = q k_\mu k^\nu$ is the radiation part of the EM tensor
in geometric optics limit, for which $k_\mu k^\mu= 0$.
In interior co-moving coordinates
\begin{equation}
ds^2=A^2c^2dt^2-B^2dR^2-R(r,t)^2(d\theta^2+sin^2\theta d\phi^2)
\end{equation}
and in the exterior radiating metric:
\begin{equation}
ds^2=a^2c^2dt_r^2+2bcdt_rdR-R(r,t_r)^2(d\theta^2+sin^2\theta d\phi^2)
\end{equation}
where $dt_r = dt - R/c$ is the retarded observer time.

Now in the interior co-moving coordinates, the luminosity of radiation is
$L = 4\pi R^2 q c$, $q = E_\mu^\nu = E_0^0 c^2$ is the energy
density of radiation, $\Gamma = dR / dl$, $U =  dR /d\tau$ and
\begin{equation}
\Gamma^2  = 1-\frac{R_{schw}}{ R} +(\frac{U}{c})^2
\end{equation}
From the boundary condition on the time like collapsing surface
$R_s = R_s(r ,t_r)$ separating
the interior metric from the exterior metric , and the POE, we
find that the proper time of the collapsing surface is given by
\begin{equation}
d\tau_s = \frac{dt_r}{1 + z_s} = dt_r(\Gamma_s + U_s / c) > 0
\end{equation}
To analyze the physical gravitational collapse process one proceeds
to solve the Einstein equations for the interior co-moving metric
with boundary conditions which connect
the interior metric to the exterior metric at the collapsing surface.
When this is done, as described by the Einstein
equations in the co-moving metric (Hernandez Jr.\& Misner, 1966,
Lindquist, Schwartz \& Misner 1965, Misner 1965, Lindquist,
1966), we find that
among the various equations associated with the collapse
process there are three proper time differential equations
which control the physical properties of a compact collapsing
and radiating physical surface. When evaluated on the
physical surface these equations are given by:
\begin{equation}
\frac{dU_s}{d\tau} = (\frac{\Gamma}{\rho+P/c^2})_s (-\frac{\partial P}{\partial R})_s
- (\frac{G(M + 4\pi R^3 (P + q ) / c^2)}{R^2})_s
\end{equation}
\begin{equation}
\frac{dM_s}{d\tau}  =  - (4\pi R^2 P c \frac{U}{c})_s - (L (\frac{U}{c} + \Gamma))_s
\end{equation}
\begin{equation}
\frac{d\Gamma_s}{d\tau}  = \frac{G}{c^4}(\frac{L}{R})_s + \frac{U_s}{c^2} (\frac{\Gamma}{\rho+ P /c^2})_s
(-\frac{\partial P}{\partial R})_s
\end{equation}

Now we have seen from Section 2 that in all frames of reference the POE requires
that $1 / (1 + z_s) > 0$ must hold in the context of the time-like
gravitational collapse of physical matter. This
POE therefore implies that a surface of infinite
red-shift cannot be dynamically formed by the time-like
collapse of physical matter. Since the POE dynamic condition
that $1 / (1 + z_s) > 0$ holds for the time-like motion
of physical matter in all frames of reference, it is also true in the
context of the co-moving frame of reference for the
equations of motion of the gravitationally collapsing
surface defined by equations (6 - 8). Hence the Principle of
Equivalence requires that equations (6 - 8) must be
dynamically constrained to obey
\begin{equation}
\frac{1}{1 + z_s} = \Gamma_s + \frac{U_s}{c}
            = (1- \frac{R_{schw}}{R} + \frac{U_s^2}{c^2})^{1/2} + \frac{U_s}{c} > 0
\end{equation}
Solving this equation for the quantity  $R_{schw} / R$  we
find this requires that the \textit{ `no trapped surface condition'}
\begin{equation}
\frac{R_{schw}}{R_s}   < 1
\end{equation}
must hold in the equation of motion for physical matter.

It follows that the POE dynamically requires, in equations of motion for
physical matter, that the \textit{ `no trapped surface condition'},  $R_{schw} / R_s < 1$ must
be maintained by the physical forces which act during the collapse process.
Hence we conclude that the Principle of Equivalence requires
that the physical processes involved in the gravitational collapse of a
physical plasma must allow it to heat up and radiate so as to allow a
high red shift Eddington limited secular equilibrium to form in a
manner which allows $R_{schw} / R_s < 1$ to be maintained.
For a physical plasma undergoing a spherically symmetric gravitational collapse,
this implies that a Magnetospheric Eternally Collapsing Object (MECO) can
dynamically form as a high red shift Eddington limited
secular equilibrium object, where both  $dU_s/d\tau = 0$
and $U_s(\tau_{edd}) = 0$  hold for the time-like motion of the surface layer
after the proper time $\tau_{edd}$. At this time the invariant
surface radius is $R_s(\tau_{edd}) = R_{s,Edd} > R(\tau_{edd})_{schw}$
whose large but finite value of the surface red shift
is given by $(1 + z_s(\tau_{edd})$.

Then for high red shift Eddington limited MECO
equation (6) with  $(U_s/c) = 0$ becomes
\begin{equation}
\frac{dU_s}{d\tau} = \frac{\Gamma_s}{(\rho + P/c^2)_s}(-\frac{\partial P}{\partial R})_s
  - \frac{G(M_s}{R_s^2}  = 0
\end{equation}
Where
\begin{equation}
M_s = (M + 4\pi R^3(P + q )/c^2)_s 
\end{equation}

Equation (11) when integrated over a closed surface
for $\tau >= \tau_{edd}$  can be solved for the net
outward flow of Eddington limited luminosity through the surface.
Since the surface $R_s(\tau_{edd})$ obeys
$R(\tau_{edd})_{schw} < R_s(\tau_{edd}) < R(\tau_{edd})_{photon}$,
then $\Gamma(\tau_{edd})_s > 0$ and the photon escape cone factor
$27 (R(\tau_{edd})_{schw}/2 R_s(\tau_{edd}))^2
(1 - R(\tau_{edd})_{schw} /R_s(\tau_{edd})))$
( $\sim (27/4)(1 + z_{edd})^2$ for $R_s \sim R_{schw}$) must be taken
into account in the calculation of Eddington limit luminosity.

When this is done one finds that the outflowing (but not all
escaping) Eddington luminosity
emitted from the surface is (see Appendix B) given by
\begin{equation}
L_{edd}(outflow)_s  =\frac{4\pi G M(\tau)_s c(1 + z_{edd})^3}
    {27 \kappa u_s^2}
\end{equation}
where $u_s=GM_s/c^2R_s$. (For simplicity, we have assumed here that the
luminosity actually escapes from the MECO surface rather than after
conveyance through an atmosphere and photosphere. The end result is the same for distant observers.)
However the luminosity $L_s$ which appears in equations (7-8) is actually
the net luminosity, which escapes through the photon radius, and is given by
$L_s = L_{edd}(escape)_s = L_{edd}(outflow) - L_{edd}(fallback) =
L_{Edd,s}-L_{Edd,s}(1-27(R_{schw}/2R_s)^2(1+z_{edd})^2$
Thus in equations 7 and 8, the $L_s$ appearing there is given by
\begin{equation}
L_s= L_{edd}(escape)_s = \frac{4\pi GM(\tau)_s c}{\kappa (1+z_{edd})}
\end{equation}
In this context from (7) we have that
\begin{equation}
\frac{dM_s}{d\tau} = -\frac{L_{edd}(escape)_s}{(1+z_{edd})^2}
       =  - \frac{4\pi G M(\tau)_s c}{\kappa c^2}  = \frac{c^2}{2G} U_{schw}
\end{equation}
which can be integrated for  $\tau > \tau_{edd}$ to give
\begin{equation}
M_s(\tau) = M_s(\tau_{edd}) \exp{((-4\pi G / \kappa c)(\tau - \tau_{edd}))}
\end{equation}
Finally, equation (8) becomes
\begin{equation}
\frac{d\Gamma_s}{d\tau}  =\frac{G}{c^4}\frac{L_{edd,s}}{R_s(\tau_{edd}}
       = \frac{4\pi G}{2\Gamma_s \kappa c}
       \frac{R(\tau)_{schw}}{R_s(\tau_{edd})}
\end{equation}
whose solution 
\begin{equation}
\Gamma_s(\tau) = \frac{1}{1 + z_s(\tau)} =
(1- \frac{R(\tau)_{schw}}{R_s(\tau)_{edd}})^{1/2}  > 0
\end{equation}
is consistent with (4) and (9).

Hence from the above we have for the high red shift MECO
solutions to the Einstein Equations that
$R_s(\tau)  = R_s(\tau_{edd})$,  $U(\tau) = U(\tau_{edd}) = 0$ and
\begin{equation}
R(\tau)_{schw} = (R(\tau_{edd})_{schw})
\exp{((-4\pi G /\kappa c)(\tau - \tau_{edd}))}
< R_s(\tau) =R_s(\tau_{edd})
\end{equation}
\begin{equation}
U(\tau)_{schw}/ c = (U(\tau_{edd})_{schw} / c)
\exp{((-4\pi G /\kappa c)(\tau - \tau_{edd})}
< U(\tau)/c = U(\tau_{edd})/c = 0
\end{equation}
where $U(\tau_{edd})_{schw} = -(R(\tau_{edd})_{schw} 4\pi G/\kappa c^2)$ and
\begin{equation}
\Gamma_s(\tau) = (1- \frac{R(\tau)_{schw}}{R_s(\tau)_{edd}})^{1/2}
= \frac{1}{1 + z_{edd})(\tau)}
= (1- \frac{R(\tau_{edd})_{schw}}{R_s(\tau_{edd})}
\exp{((-4\pi G /\kappa c)(\tau - \tau_{edd}))})^{1/2}
\end{equation}
with
\begin{equation}
1 + z(\tau)_{edd}) = (1- \frac{R(\tau_{edd})_{schw}}{R_s(\tau_{edd})}
\exp{((-4\pi G /\kappa c)(\tau - \tau_{edd}))})^{-1/2}
\end{equation}
Hence for $1+z(\tau_{edd}) \sim 10^8$ and proper time $\tau > \tau_{edd}$
it follows that distant observers see the MECO mass decay and red shift increase
with an extremely long e-folding time of $4.5\times 10^8(1+z_s)$ yr; 
$\sim  2\times 10^7 (t_{Hubble})$.

\appendix{\bf {B. Relativistic Eddington Limit Formulas}}\\
From the equation 11 of Appendix A, in the Eddington limit at a surface,
\begin{equation}
\Gamma_{edd}(-\frac{\partial P}{\partial r})_{edd}
    = (\rho+ P /c^2)_s \frac{G M_s}{R_s^2}
\end{equation}
When integrated over a closed surface for $\tau > \tau_{edd}$
equation (1) can be solved for the net
outward flow of Eddington limited luminosity through the surface.
Since the surface $R_s(\tau_{edd})$ obeys
$R(\tau_{edd})_{schw}) < R_s(\tau_{edd}) < R(\tau_{edd})_{photon}$,
then $\Gamma_s(\tau_{edd}) > 0$  and the photon escape cone factor
$27 (R(\tau_{edd})_{schw}/2 R_s(\tau_{edd}))^2
(1 - R(\tau_{edd})_{schw} /R_s(\tau_{edd})))$
must be taken into account in doing the calculation.
Then integrating this equation over a plasma shell of radius $R_s$
\begin{equation}
\int{( 4\pi R_s^2dR_s \Gamma_{edd}(-\frac{\partial P}{\partial R})_{edd}}
 =\int{( 4\pi G M_s(\rho+ \frac{P}{c^2})_s) dR}
\end{equation}
\begin{equation}
\Gamma_{edd}4 \pi \int{R_s^2 dP_{edd}}
 = 4\pi G M_s\int{(\rho+ P /c^2)_s dR}
\end{equation}
where
\begin{equation}
\Gamma_{edd} = 1 / (1 + z_{edd}).
\end{equation}
Associated with its high compactness and temperature, the MECO plasma
evolves into an optically thick pair plasma. Hence the number of protons plus
positrons will match the number of electrons in the shell. Then radiation
pressure transfers momentum to the electrons and positrons, which then
support the protons through coulomb interactions. Then if the
number of protons per cm$^2$in the shell is $N_s$, it follows that
\begin{equation}
\int{(\rho + P /c^2)_s dR} = m_pN_s,
\end{equation}
where $m_p$ is the proton rest mass. Then
\begin{equation}
\int{4\pi R_s^2 dP_{edd}} = \kappa N_s m_p L_{edd}(escape)_s / c
\end{equation}
Where $\kappa$ is MECO opacity
and $L_{edd}(escape)_s$ is the net luminosity which escapes
through the photon radius (the remainder falls back into the shell).
It is given by $L_{edd}(escape)_s = L_{edd}(outflow) - L_{edd}(fallback) =
L_{Edd,s}-L_{Edd,s}(1-27(R_{schw}/2R_s)^2(1+z_{edd})^2$
Solving equations (2-6) we have locally that
\begin{equation}
L_{edd}(outflow)_s = \frac{4\pi G M_s(\tau) c(1 + z_{edd})^3}{27 \kappa u_s^2}
\end{equation}
where $u_s = GM_s/c^2R_s$ and hence that
\begin{equation}
L_{edd}(escape)_s =  \frac{4\pi G M_s(\tau) c}{\kappa (1 + z_{edd})}
\end{equation}
which implies that the red shift at which the
Eddington limit is established is given by
\begin{equation}
1 + z_{edd} = (\frac{L_{edd}(outflow)_s 27 u_s^2 \kappa}{4\pi G M_s})^{1/3}
\end{equation}
while the net Eddington luminosity seen at infinity is
\begin{equation}
L_{Edd, \infty} = \frac{L_{edd}(escape)_s}{(1+z_{edd})^2}
=\frac{4\pi G M_s c}{\kappa (1+z)} \sim \frac{1.27 \times 10^{38} m}{1+z} ~~~ erg/s
\end{equation}

\appendix{\bf C. Relativistic particle mechanics}\\
A number of standard, but useful results for relativistic mechanics are
recapitulated here. All are based upon the energy-momentum four-vector
for a free particle in the Schwarzschild geometry of a central mass.
The magnitude of this vector, given by g$^{ij}$p$_i$p$_j$,
is m$_0$$^2$c$^2$ where m$_0$ is the rest mass of the
particle. For
a particle in an equatorial trajectory ($\theta$ = $\pi$,
p$_{\theta}$ = 0) about
an object of gravitational mass M, one obtains:
$g_{00} = 1/g^{00}= -1/g_{rr} = -g^{rr} = (1-2u)$, $g_{\phi\phi} = -r^2$, where
$u=GM/c^2r$. $p_0 = E/c$, $p^r=m_0 dr/d\tau$, $p^\phi = m_0 d\phi/d\tau$.
Here E is the particle energy as judged by a distant oberver positioned where
$u=0$. In the event $m_0=0$, some parameter, $\lambda$, must be used instead
of the proper time of a photon to describe its trajectory.
With these preliminaries, the energy-momentum equation is.
\begin{equation}
\frac{E^{2}}{(1-2 u)c^{2}}-(1-2 u){p_{r}}^{2}-\frac{{p_{\phi}}
^{2}}{{r}^{2}}={m_{0}}^{2}c^{2}
\end{equation}\\
Using $p_r = -g_{rr}p^r$ and $e = E/m_0c^2$, there follows
\begin{equation}
(\frac{dr}{d\tau})^2 = c^2(e^2-(1-2u)(1+a^2u^2))
\end{equation}
Where $a=(cp_\phi/GMm_0)$ is a dimensionless, conserved angular momentum.
For suitably small energy, bound orbits occur. Turning
points for which $dr/d\tau = 0$ can be found by examining the
effective potential, which consists of all terms to the
right of $e^2$. At minima of the effective potential we find circular
orbits for which
\begin{equation}
a^{2}=\frac{1}{u-3 u^{2}}
\end{equation}
$u=1/3$ holds at the location of an
unstable circular orbit for photons (see below).
From which we see that if $p_\phi$ is non-zero there are no trajectories
for particles with both mass and angular momentum that exit from within $u=1/3$.
Thus particles with both mass and angular momentum
can't escape from within the the photon sphere. The minimum energy
required for a circular orbit would be.
\begin{equation}
E=m_{0}c^{2}\frac{(1-2 u)}{\sqrt{(1-3 u)}}
\end{equation}
In fact, however, there is an innermost marginally stable orbit for
which the first two derivatives with respect to r or u of the
effective potential vanish. This has no Newtonian physics counterpart,
and yields the well-known results: $u=1/6$, $a^2 = 12$ and $e^2 =8/9$
for the marginally stable orbit of radius $r_{ms} = 6GM/c^2$.

For a particle beginning a radial free-fall with $a=0$
the particle energy-momentum equation becomes
\begin{equation}
dr^2 = c^2 d\tau^2(e^2-(1-2u))^{1/2}
\end{equation}
But since $p_0 = E/c = g_{00} p^0 = (1-2u)m_0cdt/d\tau$, we find
\begin{equation}
d\tau = dt(1-2u)/e
\end{equation}
Therefore, on the particle trajectory
r and t are related by
\begin{equation}
dr^2/(1-2u) = c^2dt^2 (1-2u)(1-(1-2u)/e^2)
\end{equation}
Substituting this into the radial Schwarzschild metric equation
\begin{equation}
ds^2 = (1-2u)c^2 dt^2 -dr^2/(1-2u)
\end{equation}
yields
\begin{equation}
ds^2 = (1-2u)c^2 dt^2 -(1-2u)c^2 dt^2(1-(1-2u)/e^2)
    = (1-2u)^2c^2 dt^2/e^2
\end{equation}
Unless one is prepared to argue that the infinitesimal dt can approach
infinity, this result makes it quite clear that the time-like geodesic of
a radially falling particle would become null if $u=1/2$, in
violation of the Principle of Equivalence. Of course, this
result was obvious earlier in Equation 6.
Further, a stationary observer positioned at coordinate r,
would observe the particle to have radial speed
\begin{equation}
V_r =\frac{\sqrt{g_{rr}}dr}{\sqrt{g_{00}}dt} = c(1 - \frac{(1-2u)}{e^2})^{1/2}
\end{equation}
which would yield $V_r =c$ as $u = 1/2$. For $u < 1/2$, the
Lorentz factor is $\gamma= (1-V_r^2/c^2)^{-1/2} = (1+z)e$,
where $(1+z) = (1-2u)^{-1/2}$ would be the red shift for photons emitted from
$r$.

For a particle beginning a spiral descent from $r_{ms}$ with $e=\sqrt{8/9}$,
there follows:
\begin{equation}
(\frac{dr}{d\tau})^2=c^2\frac{(6u-1)^3}{9}
\end{equation}
If observed by a stationary observer located at coordinate r, it
would be observed to move with radial speed
\begin{equation}
V_r = \frac{\sqrt{-g_{rr}}dr}{\sqrt{g_{00}}dt} = \frac{\sqrt{2}c(6u-1)^{3/2}}{4}.
\end{equation}
Again, $V_r$ approaches c as u approaches 1/2.
A distant observer would would find the angular frequency of the spiral motion
to be
\begin{equation}
\frac{1}{2\pi}\frac{d\phi}{dt} = \sqrt{9\times 12/8}(c^3/GM)u^2(1-2u)/2\pi
\sim 1.18\times 10^5 u^2(1-2u)/m ~~~Hz
\end{equation}
For a 10 M$_\odot$ GBHC ($m = 10$), this has a maximum of 437 Hz and
some interesting possibilities for generating many QPO frequencies, both
high and low. For red shifts such that $u \approx 1/2$, the spiral
frequency is  $2950/(1+z)^2$ Hz.\\
\\
{\bf Photon Trajectories:} \\
The energy-momentum equation for a particle with $m_0=0$ can be rearanged as:
\begin{equation}
(1-2u)^2(\frac{p_rGM}{p_\phi c^2})^2 = (\frac{du}{d\phi})^2 =
(\frac{GME}{p_\phi c^3})^2 - u^2(1-2u)
\end{equation}
The right member has a maximum value of 1/27 for $u=1/3$. There is an
unstable orbit with $du/d\phi = 0$ for $u=1/3$. To simply have
$du/d\phi$ be real requires $p_\phi c^3/GME < \sqrt{27}$. But
$E=(1+z)p c$, where p is the entire momentum of the photon, and
$1+z = (1-2u)^{-1/2}$ its red
shift if it escapes to be observed at a large distance. Its azimuthal
momentum component will be $p_\phi/r$. Thus its escape cone is defined by:
\begin{equation}
(\frac{p_\phi}{r p})^2 < 27u^2(1-2u)
\end{equation}

\appendix{\bf {D. Magnetosphere - Disk Interaction}}\\
The torque per unit volume of plasma in the disk threaded by
magnetic field is given by
$r \frac{B_z}{4\pi} \frac{\partial B_{\phi}}{\partial z}
\sim r \frac{B_zB_{\phi}}{4\pi H}$, where H is the disk half thickness. Thus the
rate at which angular momentum would be removed from the disk would be
\begin{equation}
\dot{m}(v_K - 2\pi\nu_s r) = r\frac{B_zB_{\phi}}{4\pi H}(4\pi H \delta r).
\end{equation}
The conventional expression for the magnetosphere radius can be obtained
with two additional assumptions: (i) that the field is fundamentally
a dipole field that is reshaped by the surface currents of the inner disk and
(ii) that $B_{\phi} = \lambda B_z (1 - 2\pi\nu_s r/v_k)$, where $\lambda$
is a dimensionless constant of order unity. This form
accounts for the obvious facts that $B_{\phi}$ should go to zero at
$r_c$, change sign there and grow in magnitude at greater distances from $r_c$.
In fact, however, we should note that
we are only describing an average $B_{\phi}$ here, because it is possible that
the field lines become overly stretched by the mismatch between
magnetospheric and Keplerian disk speeds, then break and reconnect across
the disk. This type of behavior leads to high frequency oscillations and
has been described in numerical simulations (Kato 2000).
With these assumptions we obtain
\begin{equation}
r = (\frac{\lambda \delta r}{r})^{2/7}(\frac{\mu^4}{GM\dot{m}^2})^{1/7}
\end{equation}
In order to estimate $\delta r/r$,
we choose an object for which few would quibble about it being magnetic;
namely an atoll class NS. The rate of spin is typically 400 - 500 Hz,
the co-rotation radius is $\sim 26$ km, and the maximum luminosity
for the low state is $\sim 2\times 10^{36} = GM\dot{m}/2r$ erg/s,
from which $\dot{m} = 5.5\times 10^{16}$ g/s, for $M=1.4 M_\odot$.
For a magnetic moment of $\sim 10^{27}$ gauss cm$^3$, we find
that ($\frac{\lambda \delta r}{r})^{2/7} \sim 0.3$. Thus if $\lambda \sim 1$,
then $\delta r/r \sim 0.013$; i.e., the boundary region is suitably small,
though larger than the scale height of the trailing disk. In this small region
the flow changes from co-rotation with the magnetosphere to Keplerian. When
its inner radius is inside $r_c$, its weight is not entirely supported
by centrifugal forces and it provides the `dead-weight' against the
magnetopause.

The equations needed for analysis of the data in Table 1 were developed
in previous work (Robertson \& Leiter 2002). Using units of  $10^{27}$
gauss cm$^3$ for magnetic moments, $100$ Hz for spin, $10^6$ cm for
radii, $10^{15}$ g/s for accretion rates, solar mass units,
$\lambda \delta r/r = 0.013$ and otherwise obvious notation we found the
magnetosphere radius to be:
\begin{equation}
r_m~=~8\times 10^6 {(\frac{\mu_{27}^4}{m \dot{m}_{15}^2})}^{1/7}~~~~ cm
\end{equation}
A co-rotation radius of:
\begin{equation}
r_c = 7\times 10^6{(\frac{m}{\nu_2^2})^{1/3}} ~~~cm
\end{equation}
The low state luminosity at the co-rotation radius:
\begin{equation}
L_c = 1.5 \times 10^{34} \mu_{27}^2 {\nu_2}^3 m^{-1}~~~~erg/s
\end{equation}
High state luminosity for accretion reaching the central object:
\begin{equation}
L_s = \xi \dot{m}c^2 = 1.4\times 10^{36} \xi \mu_{27}^2 \nu_2^{7/3} m^{-5/3}~~~erg/s
\end{equation}
Where $\xi \sim 1$ for MECO and $\xi = 0.14$ for NS is the
efficiency of accretion to the central surface.
We have recalculated the quiescent luminosities in the soft x-ray band from
0.5 - 10 keV using the correlations of Possenti et al. (2001) with
spin-down energy loss rate as:
\begin{equation}
L_q = \beta \dot{E} = \beta 4 \pi^2 I \nu \dot{\nu}
\end{equation}
where $I$ is the moment of inertia of the star, $\nu$ its rate of spin and
$\beta$ a multiplier that can be determined from the new
$\dot{E} - L_q$ correlation for given $\dot{E}$;
i.e., known spin and magnetic moment. (In previous work we
had used $\beta = 10^{-3}$ for all objects.)
We assume that the luminosity is that of a spinning
magnetic dipole for which $\dot{E} = 32\pi^4 \mu^2 \nu^4/3c^3$,
(Bhattacharya \& Srinivasan 1995)
where $\mu$ is the magnetic moment.
Thus the quiescent x-ray luminosity would then be given by :
\begin{equation}
L_q = \beta \times \frac{32 \pi^4 \mu^2 \nu^4}{3c^3}
= 3.8 \times 10^{33} \beta \mu_{27}^2 \nu_2^4 ~~~~~erg/s
\end{equation}

As the magnetic moment, $\mu_{27}$, enters each of the luminosity equations
it can be eliminated from ratios of these luminosities, leaving relations
involving only masses and spins. For known masses, the ratios then
yield the spins. Alternatively, if the spin is known from
burst oscillations, pulses or spectral fit determinations of $r_c$,
one only needs one measured luminosity
to enable calculation of the remaining $\mu_{27}$ and $L_q$.
For most GBHC, we
found it to be necessary to estimate the co-rotation radius from
multicolor disk fits to the thermal component of low state spectra.
The reason for this is that the luminosities are seldom available across
the whole spectral hardening transition of GBHC.
For GBHC, it is a common finding that the low state inner disk radius is much
larger than that of the marginally stable orbit (e.g. Markoff, Falcke \&
Fender 2001, $\dot{Z}$ycki, Done \& Smith 1997a,b 1998, Done \& $\dot{Z}$ycki
1999, Wilson \& Done 2001). The presence of a magnetosphere
is an obvious explanation. Given an inner disk radius at the spectral
state transition, the GBHC spin frequency follows from the Kepler
relation $2 \pi \nu_s = \sqrt{GM/r^3}$.

\appendix{ \bf {E. Observational Data}}\\
The third accreting millisecond pulsar, {\bf XTE J0929-314} has been found
(Galloway et al. 2002) with $\nu_s = 1/P = 185$ Hz and period derivative
$\dot{P} = 2.69\times 10^{-18}$, from which the magnetic field
(calculated as $3.2\times \sqrt(P\dot{P})$ is $3.9\times 10^9$ gauss.
This is typical of a Z source. Assuming a NS radius of 13 km,
the magnetic moment is $BR^3=8.5\times 10^{27}$ gauss cm$^3$.
The calculated low state limit co-rotation luminosity is $L_c=4.9\times 10^{36}$
erg/s. Approximately 40\% of this would be the luminosity in the (2 - 10 keV)
band. This yields an expected flux of $2\times 10^{-10}$ erg/cm$^2$/s for
a distance of 9 kpc. This corresponds to the knee of the published light curve
where the luminosity begins a rapid decline as the propeller becomes
active. Similar breaking behavior has been seen in Sax J1808.4-3659 and
GRO J1655-40 at propeller onset. The predicted 0.5-10 keV band luminosity
is $L_q = 1.3\times 10^{33}$ erg/s.

The second accreting millisecond pulsar {\bf XTE J1751-305} was found with a
spin of 435 Hz. (Markwardt et al. 2002) Its spectrum has been analyzed
(Miller et al. 2002). We find a hard state luminosity of
$3.5\times 10^{36}$ erg/s ($d=8$ kpc) at the start of the rapid
decline which is characteristic of the onset of the propeller
effect. We take this as an estimate of $L_c$.
From this we estimate a magnetic moment of $1.9\times 10^{27}$
gauss cm$^3$ and a quiescent luminosity of $5\times 10^{33}$ erg/s. An
upper limit on quiescent luminonosity of $1.8\times 10^{34}$ erg/s
can be set by the detections of the source in late April 2002, as
reported by Markwardt et al. (2002).

The accreting x-ray pulsar, {\bf GRO J1744-28} has long been cited for exhibiting
a propeller effect. Cui (1997) has given its spin frequency as 2.14 Hz and
a low state limit luminosity as $L_c = 1.8\times 10^{37}$ erg/s (2 - 60 keV.),
for a distance of 8 kpc. These imply a
magnetic moment of $1.3\times 10^{31}$ gauss cm$^3$ and a
magnetic field of $B=5.9\times 10^{12}$ gauss for a 13 km radius.
It spin-down energy loss rate should be $\dot{E} = 1.4\times 10^{35}$ erg/s
and its quiescent luminosity, $L_q = 3\times 10^{31}$ erg/s.
Due to its slow spin, GRO J1744-28 has a large co-rotation radius of
280 km. A mass accretion rate of $\dot{m}= 5.4\times 10^{18}$ g/s is needed
to reach $L_c$. Larger accretion rates are needed to reach the star surface,
but such rates distributed over the surface would produce luminosity in
excess of the Eddington limit. The fact that the magnetic field is strong
enough to funnel a super-Eddington flow to the poles is the likely reason for the
type II bursting behavior sometimes seen for this source. In addition to
its historical illustration of a propeller effect, this source exemplifies
the inverse correlation of spin and magnetic field strength in accreting
sources. It requires a weak field to let an accretion disk
get close enough to spin up the central object. For this reason we
expect Z sources with their stronger B fields to generally spin more
slowly than atolls.

The accreting pulsar, {\bf 4U0115+63}, with a spin of 0.276 Hz and a magnetic
field, derived from its period derivative, of $1.3\times 10^{12}$ gauss
(yielding $\mu = 2.9\times 10^{30}$ gauss cm$^3$ for a 13 km radius)
has been shown (Campana et al. 2002) to exhibit a magnetic propeller
effect with a huge luminosity interval from $L_c = 1.8\times 10^{33}$
erg/s to $L_{min} = 9.6\times 10^{35}$ erg/s. $L_c$ held steady precisely at
the calculated level for a lengthy period before luminosity began increasing.
Due to the slow spin of this star, its quiescent luminosity, if
ever observed, will be just that emanating from the surface. Its spin-down
luminosity will be much too low to be observed.

The atoll source {\bf 4U1705-44} has been the subject of a recent study
(Barret \& Olive 2002) in which a Z track has been displayed in a
color-color diagram. Observations labeled as 01 and 06 mark the end
points of a spectral state transition for which the luminosity ratio
$L_{min}/L_c = 25.6\times 10^{36}/6.9\times 10^{36} = 3.7$
can be found from their Table 2. These yield $\nu = 470$ Hz and
a magnetic moment of $\mu = 2.5\times 10^{27}$ gauss cm$^3$.
The spin-down energy loss rate is $1.2\times 10^{37}$ erg/s and
the 0.5 - 10 keV quiescent luminosity is estimated to be about
$5\times 10^{33}$ erg/s. At the apex of the Z track
(observation 12), the luminosity was $2.4\times 10^{37}$ erg/s (for
a distance of 7.4 kpc.); i.e., essentially the same as $L_{min}$.
Although 4U1705-44 has long been classified as an atoll source, it
is not surprising that it displayed the Z track in this outburst as
its 0.1 - 200 keV luminosity reached 50\% of the Eddington limit.

Considerable attention was paid to reports of a truncated accretion
disk for the GBHC, {\bf XTE J1118+480} (McClintock et al 2001) because
of the extreme interest in advective accretion flow (ADAF) models
for GBHC (Narayan, Garcia \& McClintock 1997). McClintock et al, fit
the low state spectrum to a disk blackbody plus power law model and found
that the disk inner radius would be about 35R$_{schw}$, or 720 km for
7 M$_\odot$. Using this as an estimate of the co-rotation radius we
find the spin to be 8 Hz. The the corresponding low state
luminosity of $1.2\times 10^{36}$ erg/s (for $d= 1.8$ kpc) lets us
find a magnetic moment of $10^{30}$ gauss cm$^3$. The calculated spin-down
energy loss rate is $1.5\times 10^{35}$ erg/s and the quiescent luminosity
would be about $3\times 10^{31}$ erg/s.

A rare transition to the hard state for {\bf LMC X-3} (Soria, Page \& Wu 2002,
Boyd et al. 2000) yields
an estimate of the mean low state luminosity of $L_c = 7\times 10^{36}$
erg/s and the high state luminosity in the same 2 - 10 keV band is
approximately $6\times 10^{38}$ erg/s at the end of the transition to the
soft state. Taking these as $L_c$ and $L_{min}$ permits the estimates of
spin $\nu = 16$ Hz and magnetic moment $\mu = 8.6\times 10^{29}$ gauss cm$^3$,
assuming 7 M$_\odot$. From these we calculate a quiescent luminosity of
$10^{33}$ erg/s.

\appendix{\bf{ F. Pair Plasma Photosphere Conditions}}\\
Although we used a characteristic temperature of a pair plasma
to locate the photosphere and find its temperature, essentially the
same results can be obtained in a more conventional way.
The photosphere condition is that (Kippenhahn \& Weigert 1990):
\begin{equation}
n\sigma_T l = 2/3,
\end{equation}
where $n$ is the combined number density of electrons and positrons
in equilibrium with a photon gas at temperature T,
$\sigma_T = 6.65\times 10^{-25} cm^2$ is the Thompson scattering
cross section and $l$ is a
proper length over which the pair plasma makes the transition from
opaque to transparent. Landau \& Lifshitz (1958) show that
\begin{equation}
n=\frac{8\pi}{h^3}\int_{0}^{\infty} \frac{p^2dp}{\exp{(E/kT)}+1}
\end{equation}
where p is the momentum of a particle, $E=\sqrt{p^2c^2+m_e^2c^4}$,
k is Boltzmann's constant, $h$ is Planck's constant and $m_e$, the mass of an
electron. For low temperatures such that $kT < m_ec^2$ this becomes:
\begin{equation}
n \approx 2(\frac{2 \pi m_ekT}{h^2})^{3/2} \exp{(-m_ec^2/kT)}
\end{equation}
It must be considered that the red shift may change significantly over
the length $l$, and that $(1 + z_p)$ will likely be
orders of magnitude smaller than $(1 +z_s)$. Neglecting algebraic signs,
we can differentiate equation 14 to obtain the coordinate length over
which z changes significantly as:
\begin{equation}
\delta r = \frac{R_g\delta z}{u^2(1+z)^3}
    \approx \frac{R_g}{u^2(1+z)^2}
\end{equation}
where we have taken $\delta z \approx (1+z)$. For values of z
appropriate here we take $u = 1/2$. We estimate
$l=\delta (1+z) = 4R_g/(1+z)$ and replace $(1+z)$ with $T/T_\infty$.
Substituting expressions for $l$ and $n$ into the photosphere
condition and substituting for $T_\infty$ from equation 18
of the main text, equation 1 yields a transcendental equation for T.

For a GBHC with $z=10^8$ and $m=10$, the solution is $T_p = 3.3\times 10^8 K$
and $(1+z_p)=2500$. Then using the radiation pressure balance condition
in the pair atmosphere, we find $T_s^4 = T_p^4(1+z_s)/(1+z_p)$, from
which $T_s = 4.6\times 10^9 K$. The number density of particles
at the photosphere is $n=4\times 10^{20}$ and $10^9$ times larger at
the MECO surface. Nevertheless, the radiation pressure exceeds the
pair particle pressure there by ten fold. This justifies our
use of radiation dominated pressure in the pair atmosphere.
For an AGN with $1+z =10^8$ and $m=10^7$, we obtain photosphere and
surface temperatures of $2\times 10^8 K$ and $1.4\times 10^9 K$, respectively,
and $(1+z_p) = 50000$. We note that the steep temperature dependence of the pair
density would have allowed us to find the same photosphere temperature
within a few percent for any reasonable choice of $l$ from $10^3$ to $10^6$ cm.
In the present circumstance, we find $l=4R_g/(1+z) = 2.4\times 10^4$ cm.
This illustrates the extreme curvature of spacetime as the corresponding
coordinate interval of the distant observer is $\delta r \sim 10$ cm.\\

\clearpage

\begin{figure}
\plotone{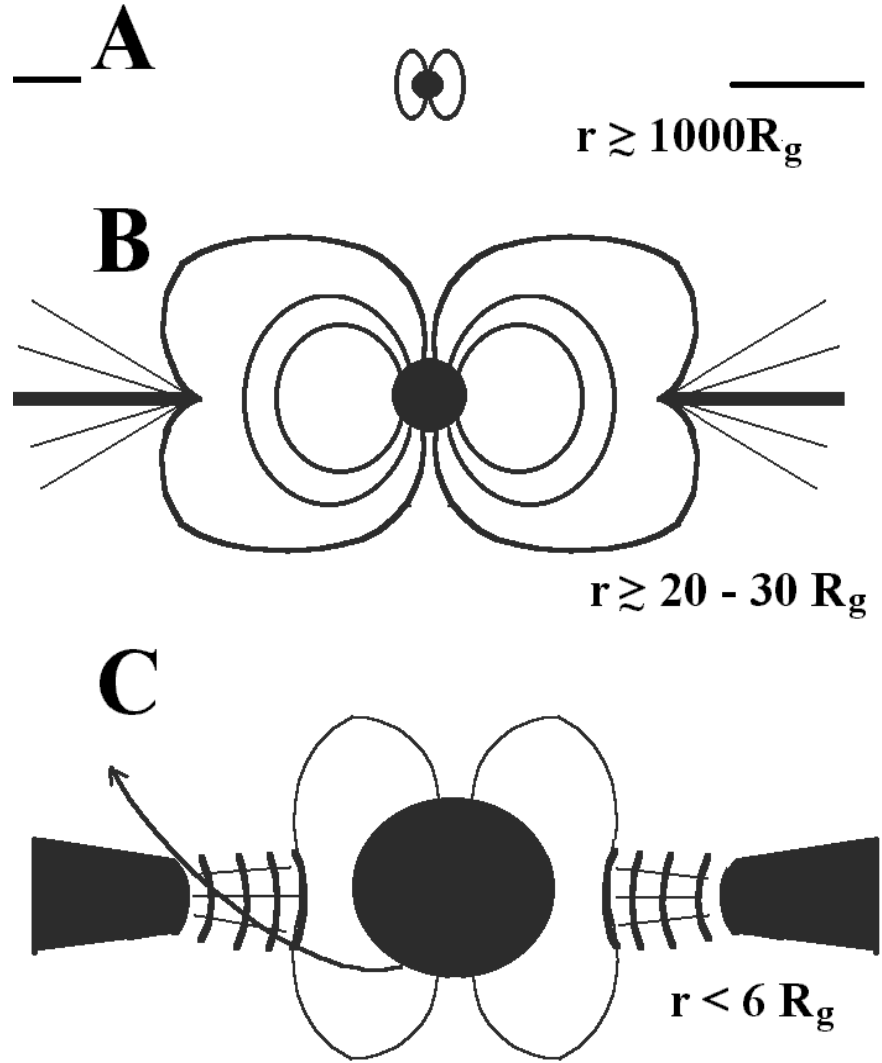}
\figcaption{MECO Spectral States:
{\bf A quiescent:}
Inner disk ablated, low accretion
rate to inner ablation radius $\sim 10^9 - 10^{10} cm$
generates optical emissions. Magnetic dipole radiation
produces hard power-law x-ray spectrum. Cooling NS or quiescent MECO emissions
may be visible.
{\bf B. Low state:}
Thin, gas pressure dominated inner disk has a radius
between the light cylinder and co-rotation radii.
Disk winds and jets are driven by
the magnetic propeller. A hard spectrum is produced as most soft
x-ray photons from the disk are Comptonized by either outflow or corona.
Outflows of electrons on open magnetic field lines produce synchrotron
radiation.
Most of the outer disk is shielded from the magnetic field of the central
object as surface currents in the inner disk change the topology of the
magnetopause.
{\bf C: High state:}
Once the inner disk is
inside the co-rotation radius, the outflow and synchrotron emissions
subside. A boundary layer of material beginning to
co-rotate with the magnetosphere may push the magnetopause to the star
surface for NS or inside $r_{ms}$ for MECO, where
a supersonic flow plunges inward until radiation
pressure stabilizes the magnetopause.
Plasma continues on to the MECO surface via interchange instabilities.
The MECO photosphere radiates a bright `ultrasoft'
thermal component. Bulk comptonization of many photons on spiral
trajectories crossing the disk produces a hard x-ray spectral tail.}
\end{figure}

\end{document}